\documentclass[letterpaper,twocolumn,10pt]{article}
\usepackage{usenix,epsfig,endnotes}

\newcommand{\ttcodefont}{\ttfamily\scriptsize}

\usepackage{color}
\usepackage{listings}
\usepackage{tabularx}
\usepackage{multirow}
\usepackage{url}

\begin{document}

\date{}

\title{\Large \bf Fast \& Flexible IO : A Compositional Approach to \\
  Storage Construction for High-Performance Devices}

\author{
{\rm Daniel G. Waddington}\\
IBM Almaden Research Center
}

\maketitle

\thispagestyle{empty}

\subsection*{Abstract}

Building storage systems has remained the domain of systems experts
for many years.  They are complex and difficult to implement.  Extreme
care is needed to ensure necessary guarantees of performance and
operational correctness.  Furthermore, because of restrictions imposed
by kernel-based designs, many legacy implementations have traded
software flexibility for performance.  Their implementation is
restricted to compiled languages such as C and assembler, and reuse
tends to be difficult or constrained.

Nevertheless, storage systems are implicitly well-suited to software
reuse and {\it compositional} software construction.  There are many
logical functions, such as block allocation, caching, partitioning,
metadata management and so forth, that are common across most variants
of storage.

In this paper, we present Comanche, an open-source project that
considers, as first-class concerns, {\it both} compositional design
and reuse, and the need for high-performance.


\section{Introduction}

Over the last decade, storage and IO system design has been
principally driven by the demand for performance.  With the advent of
a new generation of memory-based devices, the need to minimize
overhead caused by software is increasingly evident.  Nevertheless,
many of today's storage systems are hugely complex, consisting of
millions of lines of source code.  Because of this complexity,
software development and testing in this domain has become an
expensive and painstaking task.

Better component-based designs leads to improved quality (e.g., reduce
coupling and cohesion) and more effective software reuse.
Furthermore, component-based approaches can help facilitate tailoring
and reconfiguration of functionality according to changing workloads
and application requirements - that is, the development of
domain-specific solutions.

The ``panacea'' of system design is providing the duality of both
flexibility and performance.  Legacy designs, reliant on kernel-based
implementations, have achieved performance, but at a cost to software
flexibility. These designs typically rely on a top-half in userspace
and a bottom-half in the kernel.  The kernel environment is often
restricted to C and assembler, and limited to coarse-grained software
reuse.

With the increased availability of IOMMUs, more recent efforts have
focused on the use of user-level device frameworks to bypass the
kernel and ``lift'' all functionality, including device drivers, into
userspace~\cite{spdk, 7024069, Caulfield:2012:PSU:2189750.2151017,Jeong:2014:MHS:2616448.2616493,Kim:196386}.  Nevertheless, these frameworks
have had limited success in incorporating multiple programming
languages, supporting fine-grained component reuse and integrating
with legacy applications (e.g., POSIX).

This paper presents our initial exploration of achieving both
flexibility and performance in IO system design.  Our framework, known
as Comanche, leverages a userspace design strategy as the foundation
for a component-based architecture that provides fine-grained reuse
and multi-language flexibility, while retaining zero-copy and
DMA-centric optimizations where performance is paramount.  We are
currently using this framework to facilitate rapid development of
storage solutions for high-performance workloads such as genomics.

\section{Userspace Design}

A core tenet of our approach is that the complete IO stack is lifted
into userspace, including the lowest level storage device drivers
(e.g., block device).  This approach, inspired by microkernel operating system
design~\cite{Leslie2005}, relegates the kernel to control and access
control handling.  Userspace device driver placement has only recently
become viable with the advent of the IOMMU~\cite{IntelVTIO2017} and
polling-based devices~\cite{Yu:2014:OBI:2642648.2619092}.

The kernel is used to authenticate and realize the mapping of device
registers into userspace, where they can be directly accessed by a
non-privileged process (e.g., x86 Ring 3) to issue both control and
data requests.  Today, most commodity operating systems support the
necessary kernel functionality (e.g., Linux VFIO~\cite{vfio}) to
enable mapping of device registers (e.g., PCIe) to userspace memory so
that user-level processes can interact with device directly.  The main
advantages of user-level architectures are:

\begin{itemize}
\item {\it Performance} - performance is improved by eliminating the
  cost of context-switching resulting from system calls (i.e., the
  switch to IO threads) and also reducing in cache pollution.
\item {\it Resilience} - userspace deployment allows virtual memory
  protection to guard against failure proliferation.  In the kernel,
  device driver failures are often catastrophic.  In userspace,
  failure is limited to the scope of the process, which can be easily
  restarted~\cite{Sundararaman:2011:RCR:1966445.1966453}.
\item {\it Development Ecosystem} - development support in userspace
  is much greater than that of the kernel.  There are a vast array of
  library resources available to the programmer that can be easily
  integrated.  Userspace development allows different
  programming languages and compilers to be adopted.  This include more
  memory safe and provable languages ~\cite{Sigurbjarnarson:2016:PVF:3026877.3026879}.
\end{itemize}

Userspace designs have proven successful in both operating
systems~\cite{Peter:2015:AOS:2841315.2812806,
  Belay:2016:IOS:3014162.2997641,Baumann:2009:MNO:1629575.1629579,
  Colmenares:2013:TRO:2463209.2488827,Wentzlaff:2010:OSM:1807128.1807132} and storage
systems~\cite{IBM2003, Klimovic:2017:RRF:3093336.3037732, Kim:196386,
  Papagiannis:2017:IOI:3041710.3041713}. However, their success has
been limited to niche or research environments where the development
ecosystem has not been main streamed.  While, kernel-based
implementations do provide some level of API and coarse-grained
component reuse (e.g., shared device drivers), the kernel suffers from
lack of memory protection necessary to guard against ``arbitrarily''
introduced misbehaving code.  This is a fundamental problem for
software composition in the kernel.

\subsection{Software Component Model}

Our component model is inspired by Microsoft COM (Component Object
Model) ~\cite{2006essential, rogerson1997}, the component framework
that later became the foundation of the Microsoft .NET architecture.
As the basis for component software design~\cite{SGM02}, COM provides
support for interface management (in terms of syntax and semantics),
dynamic binding/composition and reference counting.

Comanche (COM-like) components are implemented as dynamically linked
libraries (DLL) that can be loaded and bound at runtime. Interfaces
are realized as C++ classes and can be inherited according to the C++
rules.  Each DLL implements a {\it factory} interface for one or more
components.  The factory interface provides methods to instantiate the
corresponding components and get hold of their base interface, {\tt
  IBase}.  {\tt IBase} is implemented by all components and provides
reference counting, typed interface querying and requests, and
dependency binding.  

The {\tt query\_interface} method takes a UUID (Universally Unique
Identifier) that defines the interface type and also designates its
syntax and semantics.  Only the interface definition, defined as a C++
class with pure virtual methods, is required to interact with a
component.  All data structures needed are defined as part of the
interface.

Components can be instantiated within components themselves.  This is
known as component {\it aggregation}.  They can also be bound to each
other by exchanging interface pointers.  This is known as component
{\it composition}. Compositional binding to other components is
supported through the {\it IBase::bind} method which itself takes an
{\it IBase} interface reference.

\subsection{Decomposing Storage Functions into Components}

The Comanche framework provides interface definitions and basic
implementations for key storage functions.  The ethos of the
framework is that different embodiments of these functions can be
implemented and easily integrated because they share a clearly defined
common interface.  Although not all of these have yet been implemented
in the current prototype, Table~\ref{table:excomp} offers example
component categories.

\begin{table}
\begin{center}

\begin{tabularx}{1.0\linewidth}{|
>{\setlength{\hsize}{.2\hsize}\raggedright\footnotesize}X|
>{\setlength{\hsize}{.7\hsize}\raggedright\arraybackslash\footnotesize}X|}\hline

  {\bf Block Devices} &
  Storage device access: NVMe, AHCI, POSIX file-based emulation \\\hline

  {\bf Allocators} &
  Space management: Block, memory (slab, heap), bitmap, tree \\\hline

  {\bf Metadata} &
  Metadata Management: On-disk, in-memory, file-based, KV-based, database \\\hline

  {\bf Index} &
  Data indexing: B-tree, R-tree, Hash table, etc. \\\hline

  {\bf Check-Pointing} & Snapshot Management: Copy-on-write, re-direct
  on write, incremental, clone/split-mirror, CDP\\\hline
  
  {\bf Persistent Memory} & PM: support for memory-centric persistence
  on NVDIMM or NVMe-backed DRAM. Fixed or paged.\\\hline

  {\bf Partitioning} &
  Fixed partitioning: GPT, virtual devices \\\hline

  {\bf Caching} &
  Cache functions: block, memory \\\hline

  {\bf RAID} &
  RAID functions: striping, mirroring, erasure coding \\\hline

  {\bf Filesystems} &
  Filesystems: traditional file-based access. \\\hline

  {\bf Operations} &
  Storage operations: replication, de-duplication, encryption, copying/DMA. \\\hline

  {\bf Tiering} &
  Auto-tiering, hierarchical storage management. \\\hline

  {\bf Network} &
  Network operations: geo-replication, sharing, RDMA, etc. \\\hline

\end{tabularx}
\caption{Example Component Categories}
\label{table:excomp}
\end{center}
\end{table}

Applications are built by composing (binding) components together.
Components can be dynamically reconfigured, providing that the appropriate
locking is put in place.

\section{Performance Opportunities}

A key motivation for breaking away from legacy POSIX and operating
system APIs for IO, is to attain a more DMA-centric capability.
Modern high-performance compute, network and storage devices rely on
DMA engines to move blocks of data without depending on CPU resources
to do so.  By using DMA, more CPU cycles can be made available for
other compute (e.g., encryption, data processing).  Many DMA engines
also provide capabilities beyond copy, such as fill.  Table~\ref{table:dma}
lists DMA transfer rates of a selection of state-of-the-art devices.

\begin{table}[h]
\begin{center}

\begin{tabularx}{1.0\linewidth}{|
    >{\setlength{\hsize}{.25\hsize}\raggedright\footnotesize}X|
    >{\setlength{\hsize}{.30\hsize}\raggedright\footnotesize}X|
    >{\setlength{\hsize}{.25\hsize}\raggedright\footnotesize}X|
    >{\setlength{\hsize}{.2\hsize}\raggedright\arraybackslash\footnotesize}X|}\hline

  Technology & Description & Max. Throughput & Latency \\\hline
  
  {\bf Intel IOAT (Crystal Beach)} &
  DRAM-to-DRAM 4K transfers &
  5963 MiB/s (QD=8) &
  0.68 $\mu$s latency (QD=1)\\\hline

  {\bf Intel Optane P4800X NVMe SSD} &
  DRAM-to-SSD 4K Random Read/Write &
  2.38 GiB/s &
  ~10 $\mu$s latency (QD=1) \\\hline  
  
  {\bf Mellanox ConnectX-4 (100GbE)} &
  NIC-to-DRAM 4K transfers &
  ~11 GiB/s &
  2 $\mu$s (1-switch, 2-hop) \\\hline

  {\bf NVIDIA GM204GL Tesla M60 GPU} &
  DRAM-to-GPU.DRAM &
  1.0 GiB/s (4KiB IO)
  7.2 GiB/s (4MiB IO) &
  15 $\mu$s \\\hline

\end{tabularx}
\caption{State-of-the-art DMA Capabilities}
\label{table:dma}
\end{center}
\end{table}

\vspace{-4mm}
\subsection{Zero-copy}

For a DMA engine to operate on memory, it must be pre-paged
and {\it pinned} so that it cannot be inadvertently remapped by the
operating system.  Pinning memory is different from {\it locking},
which can be achieved through the POSIX {\tt mlock}
API~\cite{Corbet2014}.  A page that is `locked' by the kernel dictates 
that there is always a physical mapping and that page-faults cannot
happen.  Nevertheless, the kernel may choose to remap or migrate a
page.  For the purposes of DMA, memory must be pinned and prevented
from remapping.  As of writing, the POSIX APIs do not provide any APIs
that allow an application to allocate pinned memory.  User-level
device driver frameworks, such as SPDK~\cite{spdk}, do provide APIs to
allocate contiguous pinned memory for DMA.

\begin{figure}
\vspace{5mm}
\centering
\includegraphics[width=0.8\columnwidth]{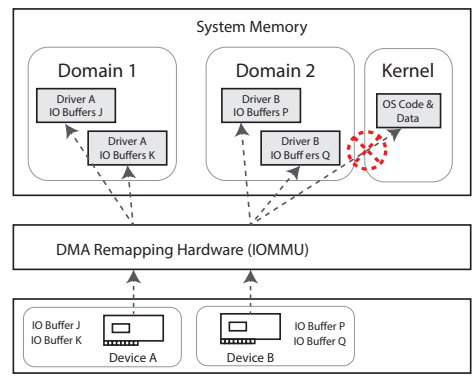}
\caption{IOMMU Protection for User-level Device Drivers}
\label{fig:iommu}
\vspace*{-4mm}
\end{figure}

Another important element of user-level device driver realization is
the integration of the IOMMU~\cite{IntelVTIO2017}.  The IOMMU provides
the ability to isolate the region of physical address space a device
is permitted to perform DMA operations on.  Without an IOMMU, any
user-level device driver would have free-reign to access any location
in system memory. This would effectively give the user-level process
root-level privilege, which is not desirable.  Thus, to protect
against arbitrary memory accesses, the IOMMU provides hardware
translation between an IO Virtual Address (IOVA) and a physical
address (see Figure~\ref{fig:iommu}). By allowing memory allocated to
the user-level process (for the purpose of IO) to be associated with a
specific device, an application can be implicitly restricted to
performing DMA with the given memory on the respective device.

It is also possible to establish {\it peer-to-peer} DMA where by an
application uses device register control (through the IOMMU) to
instigate DMA transfers directly from one device address space (e.g.,
SSD memory) to another (e.g., GPU memory).  This avoids unnecessary
copies into main memory when copying between devices.

A clear issue is the inability of the standardized POSIX API, and thus
legacy applications and libraries, to effectively leverage DMA and
zero-copy technologies.  To take advantage of these new technologies
we must introduce new APIs that allow ``DMA friendly'' memory to be
allocated from, and associated to, the logical device. Note, that
while {\tt mmap} provides a way to manage memory mapping, it does not
provide the necessary memory pinning capabilities previously
discussed.

Comanche provides DMA-compatible memory management interfaces and
presents DMA operations as a first-class concern for data
movement. For example, the block device interface inherits a memory
management interface ({\tt IZerocopy\_memory}) that allows memory to
be allocated for DMA operations with the respective device (see
Appendix A).

\subsection{Low-Latency}

\begin{figure}
\centering
\includegraphics[width=0.7\columnwidth]{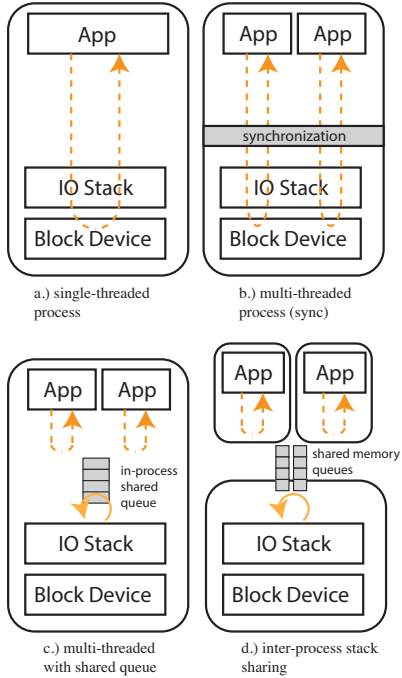}
\caption{Flexible Threading Architectures}
\label{fig:threadarch}
\vspace*{-4mm}
\end{figure}

Comanche also enables a flexible threading model.  High-performance
block devices (such as NVMe SSD) rely on polling threads.  The exact
arrangement of threads needs to be carefully considered.  For
low-latency requirements, IO stacks (i.e. composed storage functions)
can be dynamically loaded directly into the application space.  This
method eliminates the need for a system call or context switch across
threads (e.g., via a shared memory queue) since application threads
can call IO functions directly (see Figure~\ref{fig:threadarch}a).

However, a single-threaded design does not allow IO stacks (and
storage devices) to be shared across applications.  Although this
might seem overly restrictive, in some embedded and domain-specific
applications, this assumption may be quite reasonable.  Alternatively,
if an application needs to share the IO stack across multiple threads,
in potentially different applications, then some form of
synchronization must be used.  Thread synchronization (e.g., mutex
lock) or in-process shared memory queues can enable effective stack
sharing (Figure~\ref{fig:threadarch}b and ~\ref{fig:threadarch}c). The
impact of threading model can be significant. For example, typical
latency of synchronous 4K IO operations (Queue Depth 1) on an Intel Optane
P4800X device is 7$\mu$s for a single-threaded arrangement and
12$\mu$sec for queue-based sharing across threads (single IO servicing
thread).

IO stacks can also be shared across multiple processes through a
shared memory (or IPC) design (Figure~\ref{fig:threadarch}d).  In this
case, the application can issue requests to a thread-safe 
  out-bound queue, while the IO stack signals completion in the
reverse direction through a separate queue.  IO descriptors,
passed across the shared queues, are allocated by the application
and freed by the IO stack.  This means that memory management must
also be shared and synchronized accordingly (e.g., through a lock-free
descriptor queue).

Components that require polling functionality (e.g., drivers for NVMe
or RDMA) can be configured with either {\it active} threading, where
by a thread is created for each component instance, or {\it thread
  sharing} where by a single polling thread can service multiple
components (at some cost to latency).  Thread sharing is useful when
there are many devices in the system and there is a need to coalesce
polling work to reduce busy-waiting CPU cycles.  In the current
prototype, we are able to handle over 1.5M IOPS per core (NVMe 4K
random read) using a single polling thread across three NVMe devices
and shared memory service queues connecting multiple clients
(i.e. arrangement in Figure~\ref{fig:threadarch}c).

\section{Multi-Language Embedding}

We believe that the ability to use programming languages beyond C is a
crucial advantage over traditional kernel implementations.  Because
many language runtimes depend on application libraries, such as the C
runtime library, they cannot be easily executed in a kernel.  Comanche
is aimed at enabling the inclusion of components implemented in a
variety of programming languages.

\begin{figure}
\vspace{5mm}
\centering
\includegraphics[width=0.7\columnwidth]{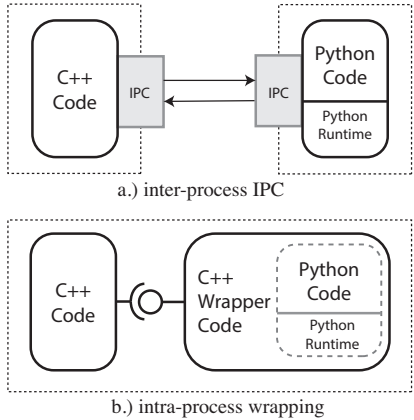}
\caption{Language Integration Approaches}
\label{fig:lang}
\end{figure}

The most straightforward approach to combining components, implemented
in different languages, is through Inter-procedural Process Calls
(IPC).  Here, separate memory spaces (i.e. processes) interact through
IPC mechanisms provided by the kernel or user-level IPC~\cite{Unrau98}
(see Figure~\ref{fig:lang}a.).  In this case, marshaling and
unmarshaling of data into the respected type system must be performed.
Frameworks such as Google Flatbuffers~\cite{flatbuffers} and
Protobuffers~\cite{protobuf} provide such capability.

The alternative is to embed components, written in different
languages, into the same process.  Because most languages do not
conform to the C++ binary interface or type system, it is necessary to
use the language's Foreign Function Interface (FFI) or equivalent, to
allow the wrapping of ``foreign code'' inside a component while
exposing a C++ interface so that it can be integrated with other
components.  In the Comanche prototype, we have demonstrated the
wrapping of Python code inside a loadable component using this method.
This could be a starting point for the integration of formally-proven
system software ~\cite{Sigurbjarnarson:2016:PVF:3026877.3026879}.

\section{Legacy Application Integration}

In the context of monolithic kernels, such as Linux, one of the key
hurdles in re-architecting the system, so as to relocate IO
functionality into userspace, is that legacy applications need to be
rewritten to take advantage of the new APIs and to facilitate direct
interaction with the IO stack.

In Comanche, we are exploring a new approach to integrating user-level
IO stacks into existing applications.  The basic concept centers
around the use of Linux FUSE (Filesystem in
Userspace)~\cite{libfuse18}.  FUSE provides the ability to forward
filesystem calls out of the kernel (from VFS) back into userspace
where they can be handled by a user-level ``service'' process.
Responses from this service are passed back down to the kernel and
forwarded back to the client application.  This scheme allows
filesystems to be implemented in userspace at some performance
penalty~\cite{fuse17}.

\begin{figure}
\centering
\includegraphics[width=0.9\columnwidth]{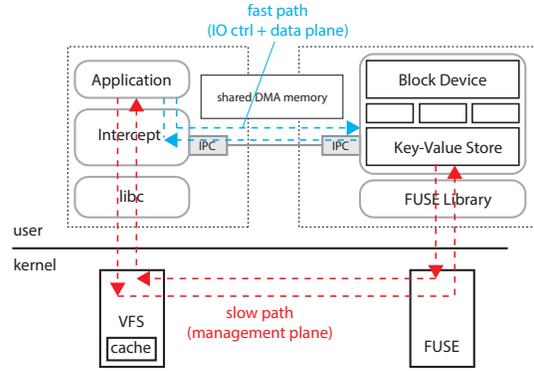}
\caption{Example FUSE Integration}
\label{fig:fuse}
\end{figure}

We propose to use FUSE to realize the ``management plane'' of Comanche
stacks - this is the slow-path.  IO control (e.g., read/write
requests) and data plane transfers continue as direct, kernel bypassed
interactions (see Figure~\ref{fig:fuse}).  This approach allows the
convenience of a filesystem abstraction to be used to manage and query
the stack.  File abstractions are mapped to underlying elements.  For
example, Figure~\ref{fig:fuse} shows the integration of a Key-Value
stack.  In this scenario, files represent keys and their contents
represent the corresponding value.  Directories may be used to define
some implicit (prefix) partitioning of the key space.  Thus, the
approach is to overlaying filesystem abstractions to some underlying
storage paradigm in order to unify the management plane.

In the case of Key-Value store, operations through the management
plane include iterating key-value pairs (i.e., directory listing),
determining attributes (e.g., value size), deleting, renaming and
copying elements.  Basic IO operations, such as read, write and fill,
are issued directly through a shared memory user-to-user IPC channel.
In order to perform zero-copy DMA, the two sides must share
DMA-compatible memory.

To support legacy (POSIX) applications, we propose to use runtime call
overloading (i.e. LD\_PRELOAD) as a means to intercept both memory
allocation and IO calls. This approach requires tracking of file
handles, correlated to those issued in the management plane, in order
to identify which calls need intercepting and forwarding through the
fast-path.  A similar approach has been successfully used by
Papagiannis et al. in their Iris
work~\cite{Papagiannis:2017:IOI:3041710.3041713}.

\section{Conclusion}

This paper has presented some initial work on Comanche, a framework
for the development of fine-grained component based storage stacks.
We are developing Comanche as a flexible approach to storage system
design and implementation, while also facilitating a more
memory-centric view of data flow through high-performance DMA capable
devices.  We ultimately hope that this framework can demonstrate
improved reuse and flexibility in support of rapid development of
domain-specific storage systems.  Currently we are exploring the use
of Comanche to construct domain-specific storage for data-intensive
genomic and data analytics workloads.

Comanche is an on-going open source project available at: {\tt https://github.com/ibm/comanche}

\onecolumn

{\footnotesize \bibliographystyle{acm}
\bibliography{references}}

\section{Appendix A. Interface Excerpts}

\begin{lstlisting}[language=C++]
  class IZerocopy_memory : public Component::IBase
  {
    ...
    virtual io_buffer_t allocate_io_buffer(size_t size, unsigned alignment, int numa_node) = 0;
    virtual status_t realloc_io_buffer(io_buffer_t io_mem, size_t size, unsigned alignment) = 0;
    virtual status_t free_io_buffer(io_buffer_t io_mem) = 0;
    ...
  }

\end{lstlisting}


\end{document}